\documentclass[journal,10pt,onecolumn]{IEEEtran}

\usepackage[left=1.5in,right=1.5in,top=1.5in,bottom=1.5in]{geometry}
\usepackage{subfig}
\usepackage{multirow}
\usepackage{longtable}

\usepackage{supertabular}
\usepackage{booktabs}

\usepackage{cleveref}

\usepackage[pdftex]{graphicx}

\usepackage{amsmath}
\usepackage{amsfonts}

\usepackage{algorithm}
\usepackage{algorithmic}

\def\showAppendix{}

\begin{document}

\title{Advanced Analytics for Connected Cars\\ Cyber Security}

\author{Matan~Levi,~\IEEEmembership{Ben-Gurion University,}
        Yair~Allouche,~\IEEEmembership{IBM Security Division,}
        and~Aryeh~Kontorovich,~\IEEEmembership{Ben-Gurion University}}

\maketitle
\thispagestyle{plain}
\pagestyle{plain}

\begin{abstract}
The vehicular connectivity revolution is fueling the automotive industry's most significant transformation seen in decades. However, as modern vehicles become more connected, they also become much more vulnerable to cyber-attacks. In this paper, a fully working machine learning approach is proposed to protect connected vehicles (fleets and individuals) against such attacks. We present a system that monitors different vehicle's interfaces (Network, CAN and OS), extracts relevant information based on configurable rules and sends it to a trained generative model to detect deviations from normal behavior. Using configurable data collector, we provide a higher level of data abstraction as the model is trained based on events instead of raw data, which has a noise-filtering effect and eliminates the need to retrain the model whenever a protocol changes. We present a new approach for detecting anomalies, tailored to the temporal nature of our domain. Adapting the hybrid approach of Gutflaish et al. (2017) to the fully temporal setting, we first train a Hidden Markov Model to learn normal vehicle behavior, and then a regression model to calibrate the likelihood threshold for anomaly. Using this architecture, our method detects sophisticated and realistic anomalies, which are missed by other existing methods monitoring the CAN bus only. We also demonstrate the superiority of adaptive thresholds over static ones. Furthermore, our approach scales efficiently from monitoring individual cars to serving large fleets. We demonstrate the competitive advantage of our model via encouraging empirical results.
\end{abstract}


%
\IEEEpeerreviewmaketitle

\section{Introduction}
In the past few years, we have been witnessing
a continual transformation of the automotive industry, whereby
new technologies are integrated into the vehicles,
changing the traditional concept as we know it and improving safety, performance and efficiency.

However, as vehicles become more connected, they also become more vulnerable to remote cyber-attacks,
as researchers have recently been pointing out.
Koscher et al. \cite{koscher2010experimental} demonstrated how,
in cases where it is possible to compromise car's internal network,
it becomes possible
to control a wide range of essential functions:
disabling the brakes, selectively braking individual wheels, stopping the engine, etc.
Following the publication of \cite{miller2015remote,HACKERSKILLJEEP},
which detailed a Jeep Cherokee being remotely hacked and stopped on a highway,
Chrysler issued a recall for 1.4 million vehicles.
Checkoway et al. \cite{checkoway2011comprehensive} presented a
remote exploitation technique using different attack vectors
(e.g., Bluetooth and cellular), which enabled
a
remote
takeover of
the vehicle as well as access to the acoustics inside the cabin and
the vehicle's location. Another
team managed to hack the wireless interface of the Tire Pressure Monitoring System and use it for eavesdropping and tracking the vehicle \cite{ishtiaq2010security}.
Further research has shown that the Passive Keyless Entry and Engine Start-Up system can also be hacked \cite{francillon2011relay}.

Protecting modern vehicles is a challenging task due to three main reasons: complexity, connectivity (large attack surface) and legacy (unsafe and outdated technologies).

In \cite{charette2009car}, it was
estimated that a premium-class automobile contains over 100 million lines of code that is executed on 70-100 Electronic Control Units (ECUs), and this number is projected
to grow to 200-300 million lines of code in the near future. This mean that such vehicles are highly complex machines with many potential vulnerabilities caused by either wrong integration or by human error.

The connectivity revolution
makes it possible for modern vehicles to become connected through a wide range of network interfaces, e.g., Wi-Fi, Cellular, Dedicated Short Range Communication (DSRC), etc. This connectivity allows manufacturers to send Over-the-Air (OTA) updates, receive diagnostic information and offer various media services. But, as the number of network interfaces increases, so does the attack surface.

Modern vehicles are controlled and monitored by tens of Electronic Control Units
that communicate over one or more internal network buses based on the {\em unsecured} Control Area Network (CAN)
protocol which has limited 1 Mbps bandwidth and a data field of 8 bytes. Since the CAN packets have no source/destination identifier,
the ECUs communicate by broadcasting packets on the CAN bus,
and each ECU decides whether the packet was intended for it.

Two main lines of defense suggested
against such attacks are message authentication/encryption and intrusion detection.
The in-vehicle network's nature makes it difficult to adopt message authentication and/or encryption.
Given the CAN protocol limitations, any cryptographic message authentication would have too weak of a key to be useful.
In \cite{nilsson2008efficient}, researchers suggested a
delayed authentication scheme that uses multiple CRC fields to compound 64 bits CBC-MAC (cipher block chaining message authentication code).
Another approach was to build a new protocol, CAN
flexible data-rate
FD,
which allows flexible data rates and longer payloads \cite{hartwich2012can}.
However, due to the fact that messages must be broadcast at a high frequency, the encryption/authentication mechanisms may lead to delays, which could impair vehicle's safety.
Moreover, authentication techniques for the in-vehicle network would not necessarily prevent attackers from remotely attacking the car and gaining access using its own network interfaces.

Hence, one is led to consider Intrusion Detection Systems (IDS).
However, rule-based IDSs are fragile and cannot cover the full range of abnormal behavior. This shortcoming is mitigated by machine learning anomaly detection systems.
One complication is that the system must
monitor not only the CAN bus traffic, but rather the different vehicle's interfaces (e.g., OS, Network and CAN).


Our contribution
consists of
a novel approach that applies cross system monitoring, adapts data abstraction and apply HMM to detect anomalies using new temporal detection technique.
Due to the connectivity revolution,
attackers can attack the vehicle without leaving any trace on the CAN bus. Therefore, monitoring the CAN bus traffic alone is not enough. We suggest to monitor the vehicle's main interfaces (e.g., OS, Network and CAN). By monitoring the different interfaces, we can detect much more sophisticated and complex attacks which combine several attack vectors from the different interfaces and does not necessarily include the CAN bus (e.g., an attacker eavesdropping using a malicious application).

Additionally,
we
propose
to train the model based on events generated from the raw data by a rule based engine which monitors the different interfaces. By preprocessing the raw data into events we provide a higher level of data abstraction, which eliminates the need to retrain the model whenever a protocol is updated and helps filtering noise. Generating events will also be effective when we will discuss the different implementation techniques in Section \ref{deployment}.


By modeling time-series data to state changes, we can detect anomalous changes in states which can indicate abnormal behavior. Therefore, We suggest to train an Hidden Markov Model as a normal behavior model Using events collected from the different interfaces.
The actual anomaly detection engine is inspired by \cite{gutflaish2017temporal}, and adapted to our fully temporal setting.

Our algorithms are trained on an abstraction of the raw data in the form of ``events''.
This abstraction eliminates the need to retrain the model each time there are protocol updates.
Additionally, it acts as a dimensionality reduction technique,
helping reduce bandwidth demand and
filter out noise.

In Section \ref{deployment} we discuss the different techniques to implement our solution. We present the hybrid approach we used, where a rule based client (data collector) is integrated into the vehicle and sends events to the backend. This technique allows us to monitor car fleets, detect cross fleet anomalies and identify correlations between different vehicles.

\section{Related Work}

\subsection{Frequency Based Techniques}
Several researches suggested using the CAN packet frequency to detect abnormal activity on the CAN bus.
Due to the limited computational power in vehicles,
Song et al. \cite{song2016intrusion} proposed a lightweight
intrusion detection algorithm based on the fact that each message ID has a regular frequency,
and when attackers inject messages into the CAN bus, the message frequency changes abruptly.
In \cite{cho2016fingerprinting}, Cho and Shin likewise
relied on
the fact that most in-vehicle network messages are periodic and broadcast over CAN
and suggested exploiting the time intervals of these periodic messages as ECU fingerprints.
These methods are mainly effective for periodic messages, so an
attacker who injects messages aperiodically may go undetected.
Moreover, when the ECU is itself the source of the malicious packets IDs,
attacks may go undetected.

\subsection{Statistical Techniques}
Another
approach to anomaly detection
is to use statistical tools to construct a ``normal''
baseline and then to identify deviations from the norm.
One idea is an entropy-based model \cite{muter2011entropy,marchetti2016evaluation}.
researchers proposed entropy-based anomaly detection technique.
In \cite{muter2011entropy},
the entropy
of the in-vehicle network
was taken as the normal behavior baseline.
The basic intuition is that due to
the clear
and restrictive specification of the in-vehicle traffic,
the entropy is relatively low and therefore, attacks
(e.g., changing packet payload, packet injection)
would cause the entropy to increase.
Marchetti et al.
\cite{marchetti2016evaluation}
carried out an
extensive experimental evaluation using hours of real CAN data.
Several
limitations of entropy-based approaches were noted.
In particular, in order to detect low-volume attacks,
one must build an anomaly detector for each message ID.
Han et al. \cite{han2015statistical} divided the data into four categories
(Engine, Fuel, Gear and Wheel) and used one-way ANOVA test to identify abnormal activity.

\subsection{Machine Learning Techniques}
Much work was done in the field of anomaly detection using
machine learning techniques \cite{chandola2009anomaly}.
Given the scope of this paper, we restrict our attention
to anomaly detection in vehicles and related applications.

\subsubsection{Classification Based Techniques}
In the automotive context, Theissler \cite{theissler2014anomaly}
proposed using a one-class SVM with the radial basis function (RBF)
kernel to learn the baseline normal behavior, and classify deviations as anomalies.
The resulting classifier is applicable to sequences of events, but does not detect point anomalies.

\subsubsection{Clustering Based Techniques}
In \cite{li2016anomaly}, Li et al. suggested to use
Gaussian Mixture Model (GMM)-based clustering
to detect flights with abnormal patterns.
The
clusters are then Characterized using their temporal distribution.

\subsubsection{Deep Learning Techniques}
Several articles proposed anomaly detection using deep learning in the automotive field
\cite{kang2016intrusion, taylor2016anomaly}.
In \cite{kang2016intrusion}, Kang and Kang suggested an intrusion detection system (IDS)
for in-vehicle networks based on Deep Neural Networks (DNN).
Unsupervised Deep belief networks (DBN) were used
to initialize the DNN parameters
as a preprocessing stage.
The data set was created using a packet generator
and anomalies were injected by manipulating packets and adding some Gaussian noise.
Another deep learning technique, proposed by Taylor et al. \cite{taylor2016anomaly},
used Long Short-Term Memory (LSTM) recurrent neural network (RNN) to detect attacks on the CAN bus.
This approach works with raw data, unlike the reduced-data abstraction proposed herein.

\subsubsection{Sequential Techniques}
Given
the nature of our data, time-series anomaly detection techniques
seem like a natural approach.
The
Hidden Markov Model (HMM) is a popular and powerful tool for modeling and analyzing time-series data.
HMMs have been widely used in different
applications,
such as gene prediction \cite{stanke2003gene}, protein structure prediction \cite{de2007hidden}, speech recognition and weather forecasting. 
A fair amount of work was also done in time-series anomaly detection in various research areas.

A Hidden Markov Model was used for monitoring patients' health conditions,
predicting future clinical episodes and initiating
alerting \cite{jiang2016intelligent, forkan2017peace}. 
An HMM was
also applied to host-based anomaly detection.
Warrender et al. \cite{warrender1999detecting} compared the performance of four methods for
modeling normal program behavior and detecting intrusions based on system calls.
They used several techniques and showed that HMM achieved the best accuracy on average.
In another paper, Hu et al. \cite{hu2009simple} suggested an efficient HMM training scheme for system-call-based anomaly intrusion detection.

A previous work, related to our approach, is \cite{narayanan2016obd_securealert}, which proposed an anomaly detection method based on HMM for the CAN bus raw data. However, this method was only tested against sudden decrease/increase in speed and RPM anomalies.

Berlin et al. \cite{berlin2016poster} introduced the idea of a Security Information and Event Management
System (SIEM) for connected vehicles located in the backend, but did not specify any concrete implementation.
We will discuss the different deployment techniques for our solution, their tradeoffs and the architecture we used in Section \ref{deployment}.

\section{Proposed Detection Technique}
Having trained a Hidden Markov Model,
it remains to specify how to identify anomalous events
based on the model likelihood score.
Most commonly,
a static threshold is used (either for  single observation or for sequences),
and scores crossing the threshold are flagged as anomalous.
However, a static threshold can be inaccurate in many applications such as
temporal data with time- and history- sensitive characteristics.

We propose a different approach for identifying anomalous events using additional regression model,
based on work done in \cite{gutflaish2017temporal} on temporal anomaly detection in databases accesses.
The training set is divided into two parts $P_1,P_2$. The first part, $P_1$, will be used to train the HMM model, while the second part, $P_2$, will be used to build a regressor that will predict the log-Likelihood
for time interval $t$.
After the Hidden Markov Model was generated, we will build the training set for the regression from $P_2$ as described in Algorithm \ref{BuildReg}.

\begin{algorithm}
\caption{BuildRegressor}\label{BuildReg}
\renewcommand{\algorithmicrequire}{\textbf{Input:}}
\renewcommand{\algorithmicensure}{\textbf{Output:}}

\begin{algorithmic}[1]
  \REQUIRE HMM model $H$,
  training set $P_2=\{ (O^1_1,\ldots,O^1_{n_1}), \ldots, (O^k_1,\ldots,O^k_{n_k}) \}$
\ENSURE Regression model $\hat{w}$
\FOR{$j \gets 1$ to $k$}
\FOR{$i \gets 1$ to $n_j$}
	\STATE $LL_{ij} \gets getLogLikelihood(H, (O^j_1,\ldots,O^j_i))$ 	
	\STATE $v_{ij}$  is a temporal feature vector created for the $i$th observation in the $j$th
        drive (e.g., time since drive start, consecutive arrival times, etc.)
\ENDFOR
\ENDFOR
\STATE $\hat{w}$ is built using least squares optimization problem: 
\[ \hat{w} \gets \operatorname{arg\,min}_{w} \sum\limits_{j=1}^k\sum\limits_{i=1}^{n_j}(\langle w, v_{ij} \rangle - LL_{ij})^2 \]
\RETURN $\hat{w}$
\end{algorithmic}
\end{algorithm}

Upon receiving new event, we compare the event log-likelihood computed from the HMM model to the the regression model predicted log-likelihood. When observing an abnormal value with respect to the regression model $\hat{w}$, the system can issue an alert.

\section{Experiments}

\subsection{Datasets}
In order to collect large amount of data and simulate car fleets, we have built a small simulation for
connected vehicles. The simulator is based on the well known traffic simulator SUMO (Simulation of Urban MObility) \cite{SUMO2012} which is widely used in different researches and projects. On top of the simulations generated by SUMO, we have integrated a data collector for each vehicle. The data sent to the backend consists of events that are collected and transmitted from each vehicle. As discussed before, the use of events instead of raw data adds a layer of abstraction that helps us filter noise and avoid protocol changes issues.

\textbf{Event Definition:} An event is a well-defined occurrence in the vehicle, generated by one of the on-board security systems (in charge of collecting different sensors' data and processing them into events). Each event consists of event type (each type is associated with unique ID) and attributes.
Event types include
login, logout, door open, door close, file access, running app, install app, app update, USB inserted,
network usage, etc.
Each event contains different attributes. A {\em file access} event contains attributes such as file type
(root, protected and public), action (read, write and execute), etc.
An
{\em install app} event contains source/destination packets, target IP, etc.
The full list of events and attributes is described in Appendix \ref{events}\ifdefined\showAppendix .\fi
\ifdefined\hideAppendix
which could be found online.
\fi

\textbf{Story Definition:} drive can be described as a sequence of events sent from the vehicle to the backend. We define a {\em story} as sequence of events that together describe a scenario. A {\em drive} is therefore
composed of stories.

In order to enrich the simulator and adapt it to connected vehicles, we have added the capability to insert stories that describe communication of connected vehicles ---
new vehicles will login as the drive starts and logout as drive ends.
Therefore, we added login and logout stories.
Connected vehicles will also install and run various application in the background such as weather,
GPS and music.
Thus, we added stories such as {\em installing application},
{\em playing music (e.g., from stream, USB and phone)}, {\em GPS access}, {\em open flows (e.g., weather)},
{\em open ports}, etc.
Another fundamental feature of connected vehicles is the ability to download road maps while driving;
therefore a {\em download map} story was added.
Moreover, since connected vehicles will need frequent firmware updates,
Over The Air updates (OTA) and USB firmware updates stories were added.
Appendix \ref{Stories} 
\ifdefined\hideAppendix
, which could be found online,
\fi
contains the full and detailed list of stories.

We have simulated a group of vehicles which carried out more than 4000 drives around the city, where each drive consists of events occurred in the vehicle and sent to the backend.

\subsection{Adding Noise}
As we aim to test our implementation in as much realistic environment as possible, noise was inserted into the data. 
We have randomly (yet in a logically consistent way) injected network usage, open flows (weather and GPS applications communicating with servers), connected devices (Bluetooth communication, USB successful/unsuccessful insertions), open/closed ports, different file accesses, drive cancellation (the driver entered the
vehicle but exits before he starts driving, perhaps because he forgot something outside)
playing music during drive from USB/Bluetooth/Stream, etc.

\subsection{Data Transformation}
We tested two different transformation techniques for converting events to HMM suitable training data.

\subsubsection{EventID Transformation}
As a baseline, and in order to examine whether rest of the features improves/reduces the algorithm performance,
the model is trained based only on the event IDs.
\subsubsection{Discrete Transformation}
This method takes the different event attributes and transforms the event into a discrete feature vector using configurable "buckets" (users will be able to define the buckets size and limits). The main reason for defining an upper limit instead of finding the max value in the data set is that it is possible for an attribute to reach an extremely high maximum value, while most values are much lower;
this could cause the majority of the values to be associated with small group of buckets,
and make the feature redundant.

For this experiments, we used the event ID, velocity, File Access options, open flows, vendor type, etc. Velocity was split into 5 buckets (0-5, 5-10, 10-20, 20-50, 50+). Open flows was split into 3 buckets (low, medium, high).
File accesses were split according to the access type (read, write and exe) and the file type (public, protected and root), vendor type was split into known (white list) and unknown, etc.

\subsection{Attack Scenarios} \label{attacks}
As described above, each drive could be described as a sequence of events sent from the vehicle to the backend. 
An unknown sequence of events (such as missing events in a known sequence,
unrealistic order of events or even known sequence of events with
attributes values that do not jibe with normal behavior) could be a sign of suspicious activity.
We can now describe several different types of attack scenarios which could be tested in our the model.

\begin{itemize}
\item \textbf{Out of order scenarios:} testing out of order events that could indicate an abnormal behavior. A simple example is the "car entry" story:

\begin{center}\textit{\{car unlock, door open, door close, seat belt on, \textbf{alarm off}, \textbf{ignition}\}}\end{center}

Since ignition event cannot appear before we disable alarm, the following sequence of events is not possible:

\begin{center}\textit{\{car unlock, door open, door close, \textbf{ignition}, seat belt on, \textbf{alarm off}\}}\end{center}

\item \textbf{USB firmware update attack:} A more interesting example is a real exploit researchers used to bypass the USB update key exchange mechanism. Cars firmware can be updated using USB (as was done after the Chrysler's hacking and recall \cite{miller2015remote, HACKERSKILLJEEP}) which contains an authentication mechanism.
  Researchers \cite{miller2015remote} found out that if one takes the original USB,
  waits for the authentication process to finish,
  extracts the original USB and replaces it with a malicious one,
  the malicious USB is authenticated and can access the car's systems.
  This vulnerability was not known before, and can cause severe damage. 
The process of normal behavior can be described as a story:
\begin{center}\textit{\{USB insert, Authentication Process, Running File From USB, File Access,	USB Extract\}}\end{center}

The following sequence of events can suggest that something is wrong and our system can detect such behavior:

\begin{center}\textit{\{USB insert, Authentication Process, \textbf{USB removed}, \textbf{USB inserted}, Running File From USB, File Access,	USB Extract\}}\end{center}

\item \textbf{Communication with unknown vendor:} one of the fundamental concepts of connected cars is the ability to download road maps while driving. This can be very helpful when the vehicle's sensors which are responsible of monitoring the surrounding fail to do so. The connected cars will communicate with known vendors while downloading the maps.
An attack could be communication with unknown vendor (e.g., malware inside the system trying to disguise communication with C\&C server), followed by abnormal network communication and/or unknown processes running in the background.

\item \textbf{OTA malicious updates:} as the USB firmware updates, OTA will also be available in connected cars. A malicious update caused by attackers could cause the installation of new applications, abnormal network communication, etc.
\item \textbf{Malicious application installation:} there is wide variety of ways to make a user install a malicious application. One of those ways is to upload a benign version to the store, and after the application is installed, it updates itself with the malicious content and/or tricks the user into granting it with high privileges. This could change the application normal behavior and cause sudden drift in file accesses, attempt to access OS or CAN infrastructure, abnormal network communication, etc.
\end{itemize}

\subsection{HMM Model Generation}
We have built various HMM models with different number of hidden states.
Then, we calculated each model's log-likelihood on a test set with $K=3$ folds cross validation. 
After comparing the different models, our algorithm chose the model with the best log-likelihood.

\subsection{Tests}
In the following experiments, our focus was on two main goals:
\begin{itemize}
  \item Test the developed algorithms and system architecture.
  \item Analyze the dynamic temporal threshold performances.
\end{itemize}
We have tested HMM models with different number of hidden states both for the \textit{EventID transformation} and the \textit{Discrete transformation}. Models were tested against test set with 1000 different benign and anomalous drives. In order to generate anomalous drives, we have injected multiple variations of the anomalies types described in \ref{attacks} into benign drives.

The experiments were conducted in both offline and online modes: \textbf{full drives} (offline) and \textbf{drive prefixes} (online). \subsubsection*{Offline mode} When testing full drives, the system waits until the driver has stopped the vehicle and logged out, then it pulls the full sequence of events occurred during the drive (and were stored in the backend) and test it against the vehicle's HMM and regression models. \subsubsection*{Online mode} In drive prefixes mode, the system tests each new event as soon as it is sent from the vehicle, and alerts when the accumulated prefix of events is identified as anomalous.

To achieve the second goal and evaluate the dynamic temporal threshold performances, we performed the same tests both on static thresholds and on our dynamic temporal threshold.

\begin{figure}[!t]
    \centering
    \begin{minipage}{0.49\textwidth}
        \centering
        \includegraphics[width=0.9\textwidth]{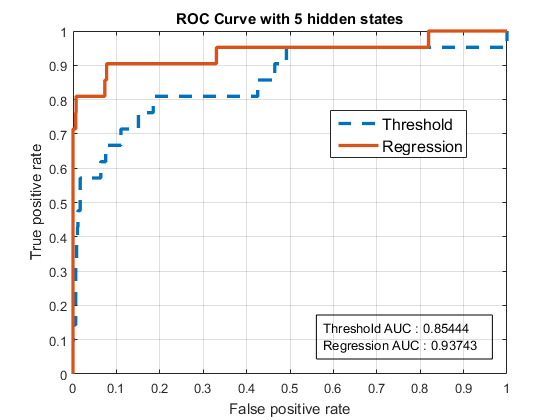}
        \caption{eventID transformation using 5 hidden states}
        \label{fig1}
    \end{minipage}\hfill
    \begin{minipage}{0.49\textwidth}
        \centering
        \includegraphics[width=0.9\textwidth]{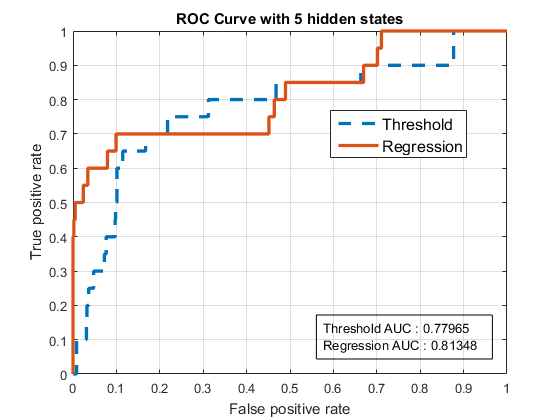}
        \caption{Discrete transformation using 5 hidden states}
        \label{fig2}
    \end{minipage}
\end{figure}

\begin{figure}[!t]
    \centering
    \begin{minipage}{0.49\textwidth}
        \centering
        \includegraphics[width=0.9\textwidth]{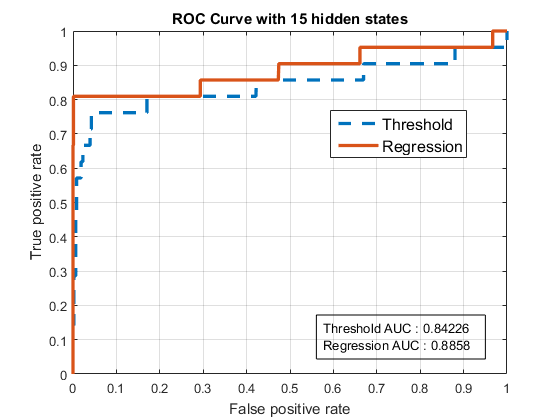}
        \caption{eventID transformation using 15 hidden states}
        \label{fig3}
    \end{minipage}\hfill
    \begin{minipage}{0.49\textwidth}
        \centering
        \includegraphics[width=0.9\textwidth]{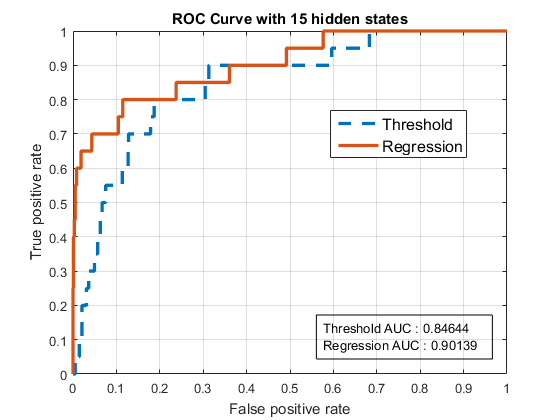}
        \caption{Discrete transformation using 15 hidden states}
        \label{fig4}
    \end{minipage}
\end{figure}

\begin{figure}[!t]
    \centering
    \begin{minipage}{0.49\textwidth}
        \centering
        \includegraphics[width=0.9\textwidth]{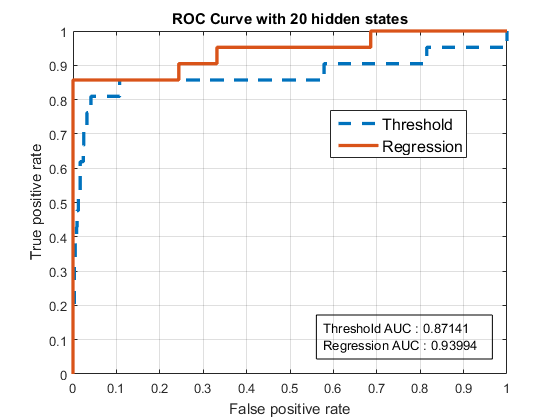}
        \caption{eventID transformation using 20 hidden states}
        \label{fig5}
    \end{minipage}\hfill
    \begin{minipage}{0.49\textwidth}
        \centering
        \includegraphics[width=0.9\textwidth]{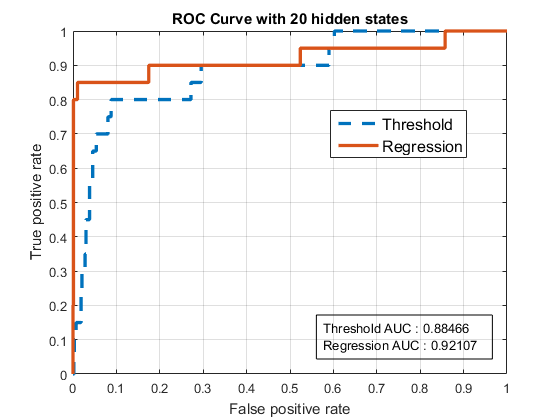} 
        \caption{Discrete transformation using 20 hidden states}
        \label{fig6}
    \end{minipage}
\end{figure}

\begin{figure}[!t]
    \centering
    \begin{minipage}{0.49\textwidth}
        \centering
        \includegraphics[width=0.9\textwidth]{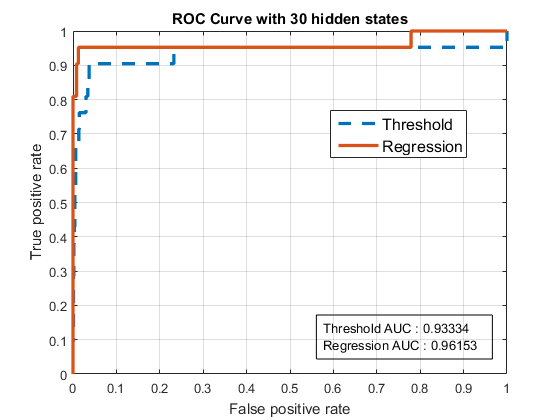}
        \caption{EventID transformation using 30 hidden states}
        \label{fig7}
    \end{minipage}\hfill
    \begin{minipage}{0.49\textwidth}
        \centering
        \includegraphics[width=0.9\textwidth]{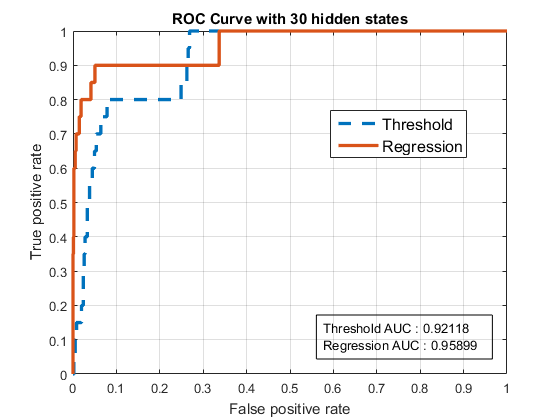} 
        \caption{Discrete transformation using 30 hidden states}
        \label{fig8}
    \end{minipage}
\end{figure}

Since the ROC-AUC and F-Measure are generally insensitive to the imbalance nature of the data, they were chosen to compare the results between different models with different hidden states, and between the two transformation techniques. ROC curves are presented in \Cref{fig1,fig2,fig3,fig4,fig5,fig6,fig7,fig8} and summarized in Table \ref{table:2}. F-Measure results are presented in Figures 9, 10 and summarized in Tables \ref{table:4}, \ref{table:5}. 

\begin{table}[!t]
\centering
\caption{AUC results under given transformation technique, detection method and number of hidden states of the HMM}
\begin{tabular}{| c c | c c c c |} 
 \hline
 \multicolumn{2}{|c|}{} & \multicolumn{4}{c|}{Hidden States Number} \\
 \cline{3-6}
   Transformation & Detection technique & 5 & 15 & 20 & 30 \\ [0.5ex] 
 \hline\hline
 EventID & Threshold & 0.85 & 0.84 & 0.87 & 0.93 \\ [0.5ex]
 \textbf{EventID}  & \textbf{Regression} & \textbf{0.94} & \textbf{0.89} & \textbf{0.94} & \textbf{0.96} \\ [0.5ex]
 Discrete & Threshold & 0.78 & 0.84 & 0.88 & 0.92 \\ [0.5ex]
 \textbf{Discrete} & \textbf{Regression} & \textbf{0.81} & \textbf{0.9} & \textbf{0.92} & \textbf{0.96} \\ [0.5ex]
 \hline
\end{tabular}
\label{table:2}
\end{table}

\begin{figure}[!t]
	\label{event_id_fmeasure}
    \centering
    \begin{minipage}{0.5\textwidth}
        \centering
        \includegraphics[width=0.9\textwidth]{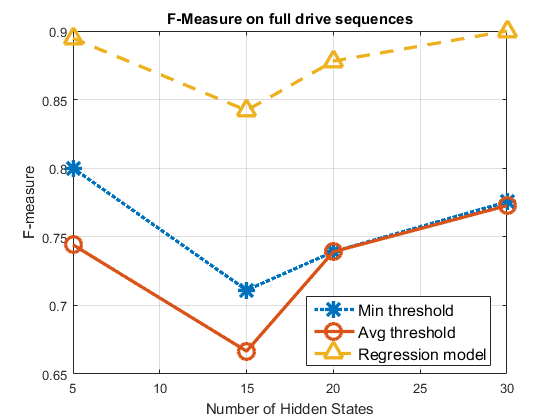}
        \subfloat{Offline comparison}
    \end{minipage}\hfill
    \begin{minipage}{0.5\textwidth}
        \centering
        \includegraphics[width=0.9\textwidth]{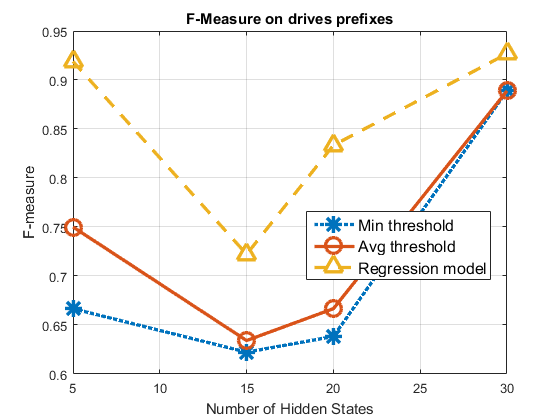}
        \subfloat{Online comparison}
    \end{minipage}
    \caption{EventID transformation F-Measure results}
\end{figure}

\begin{figure}[!t]
    \label{discrete_fmeasure}
    \centering
    \begin{minipage}{0.5\textwidth}
        \centering
        \includegraphics[width=0.9\textwidth]{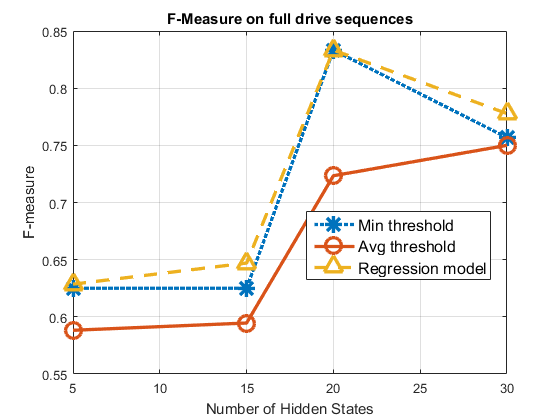}
        \subfloat{Offline comparison}
    \end{minipage}\hfill
    \begin{minipage}{0.5\textwidth}
        \centering
        \includegraphics[width=0.9\textwidth]{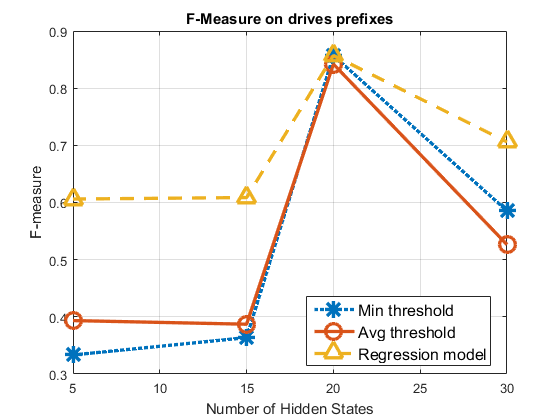}
        \subfloat{Online comparison}
    \end{minipage}
    \caption{Discrete transformation F-Measure results}
\end{figure}

\begin{table}[!t]
\centering
\caption{F-Measure results with EventID transformation}
\begin{tabular}{| c c | c c c c |} 
 \hline
 \multicolumn{2}{|c|}{} & \multicolumn{4}{c|}{Hidden States Number} \\
 \cline{3-6}
   Offline/Online & Detection technique & 5 & 15 & 20 & 30 \\ [0.5ex] 
 \hline\hline
 Offline & Avg. Threshold & 0.74 & 0.67 & 0.74 & 0.77 \\ [0.5ex]
 Offline & Min. Threshold & 0.8 & 0.71 & 0.74 & 0.78 \\ [0.5ex]
 \textbf{Offline}  & \textbf{Regression} & \textbf{0.89} & \textbf{0.84} & \textbf{0.88} & \textbf{0.9} \\
 \hline\hline
 Online & Avg. Threshold & 0.75 & 0.63 & 0.67 & 0.88 \\ [0.5ex]
 Online & Min. Threshold & 0.67 & 0.62 & 0.64 & 0.88 \\ [0.5ex]
 \textbf{Online} & \textbf{Regression} & \textbf{0.92} & \textbf{0.72} & \textbf{0.83} & \textbf{0.93} \\ [0.5ex]
 \hline
\end{tabular}
\label{table:4}
\end{table}

\begin{table}[!t]
\centering
\caption{F-Measure results with Discrete transformation}
\begin{tabular}{| c c | c c c c |} 
 \hline
 \multicolumn{2}{|c|}{} & \multicolumn{4}{c|}{Hidden States Number} \\
 \cline{3-6}
   Offline/Online & Detection technique & 5 & 15 & 20 & 30 \\ [0.5ex] 
 \hline\hline
 Offline & Avg. Threshold & 0.58 & 0.59 & 0.72 & 0.75 \\ [0.5ex]
 Offline & Min. Threshold & 0.62 & 0.62 & 0.83 & 0.76 \\ [0.5ex]
 \textbf{Offline}  & \textbf{Regression} & \textbf{0.63} & \textbf{0.65} & \textbf{0.83} & \textbf{0.78} \\ [0.5ex]
  \hline\hline
 Online & Avg. Threshold & 0.39 & 0.39 & 0.84 & 0.53 \\ [0.5ex]
 Online & Min. Threshold & 0.33 & 0.36 & 0.86 & 0.59 \\ [0.5ex]
 \textbf{Online} & \textbf{Regression} & \textbf{0.61} & \textbf{0.61} & \textbf{0.86} & \textbf{0.71} \\ [0.5ex]
 \hline
\end{tabular}
\label{table:5}
\end{table}

\section{Results}
As it can be observed from the graphs, both transformation techniques reached high AUC of 0.96. The \textit{eventID transformation} reached F-Measure higher than 0.9 both in offline and online mode. The \textit{Discrete transformation} reached F-Measure of 0.83 in offline mode, and 0.86 in online mode.

Although the F-Measure is higher for the \textit{eventID transformation}, the \textit{Discrete transformation} advantage lays with its ability to identify anomalies that could not be identified with the \textit{eventID transformation} such as: application accessing files not permitted for it, unusual behavior in different contexts (events that are legit only in specific speeds/flows/files context), etc.

Best results were achieved indisputably by using the regression model temporal threshold. The regression model was superior to any static threshold in all tests, detection modes and model measurement methods.

\begin{figure}[!t]
\centering
\includegraphics[width=\linewidth]{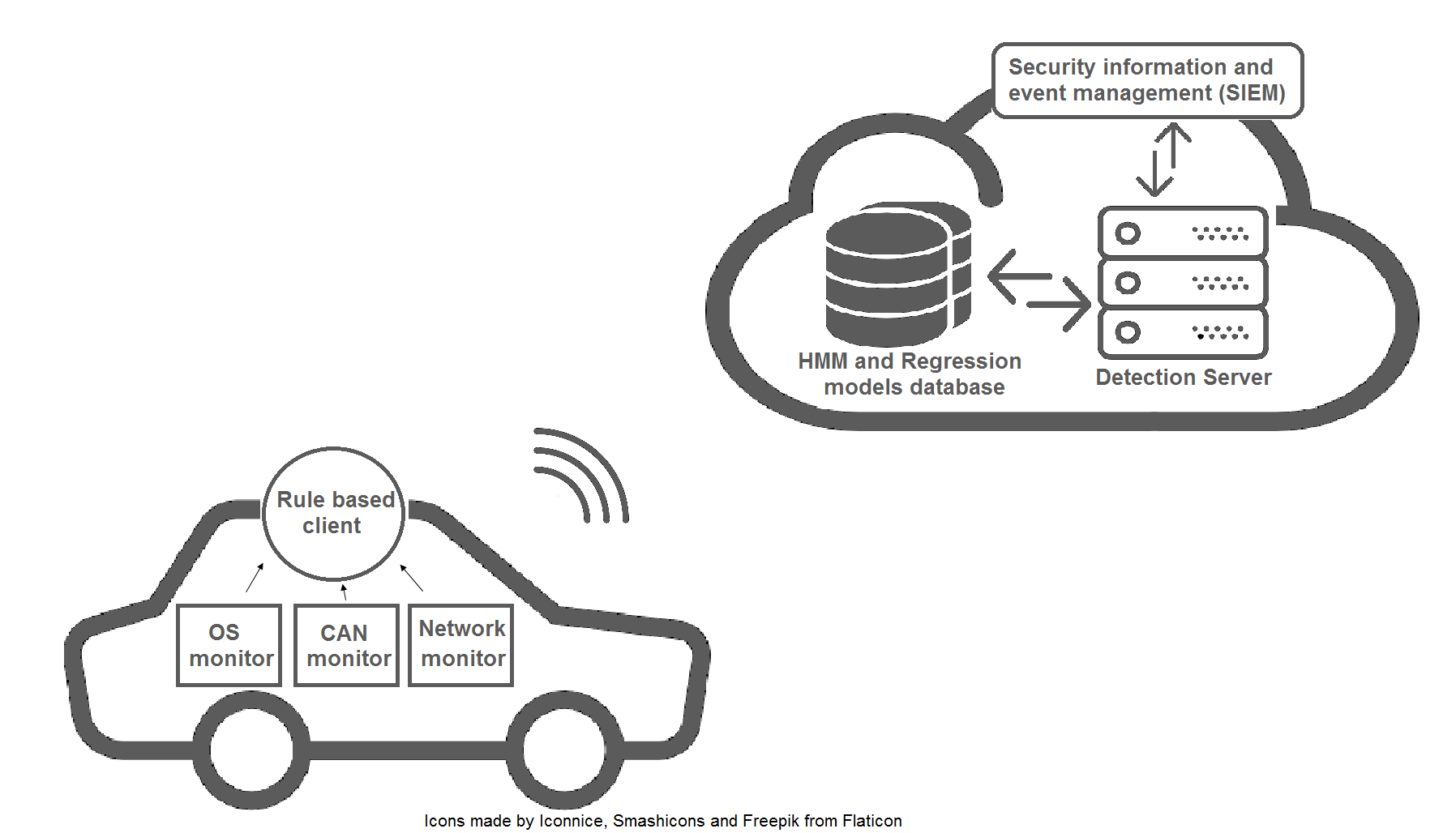}
\caption{High Level System Architecture}
\label{arch}
\end{figure}

\section{Deployment} \label{deployment}

There are two main approaches for deploying our solution: on-board or on the backend. Those approaches present a
trade-off
between
several aspects including detection latency, detection scale, footprint (resources) and bandwidth consumption.

The on-board architecture will provide real time anomaly detection due to the fast communication between the data collector and the detection engine. However, this method
is resource-intensive
on the edge units. Furthermore, it is far more challenging to serve car fleets and detect cross fleet anomalies using this architecture. On the other hand, in the backend architecture, serving car fleets and detecting cross fleet anomalies is much easier, but there is certain latency in the data transmitted from vehicles to the cloud. Moreover, this approach requires enormous bandwidth.

Our solution had to overcome two major challenges:
\begin{enumerate}
\item Apply real time anomaly detection using HMM using
  overcoming the low computational power and limited memory resources.
\item
  Detect anomalies in real time for a car fleet.
\end{enumerate}

Therefore, we chose to use a hybrid approach (known approach in the industry), where a rule based client (the data collector) is integrated into the vehicle and sends events to the backend. However, as vehicles evolve,
computational power and memory resources limitations will not be an obstacle anymore and this
solution could be integrated fully inside the vehicle.

In order to address those challenges we suggest a hybrid platform with HMM anomaly detection mechanism: 

By moving the detection mechanism to the cloud, we can overcome the low computational power and limited memory resources of the vehicle. However, transmitting raw CAN/OS/Network packets to the cloud will require enormous bandwidth, high transmission rate and can overwhelm the backend. Therefore, we suggest to use a light-weight component that could be integrated into the vehicle and will be able to extract important data based on configurable rules, and transmit it back to the backend with all the selected features, which means our data will not be raw packets, rather it would be in a higher level of abstraction and include only relevant data extracted from applications, network traffic, chosen sensors, CAN bus, etc.

The backend will store Hidden Markov Models learned for a large number of vehicles. By using a cloud-based platform we can serve car fleets and possibly apply advanced analytics for identifying correlations between vehicles and identify in-progress attacks on vehicles using anomalies detected from other vehicles.
Using this approach, we can either train an Hidden Markov Model for each car and store it in a distributed database, or build a model for groups of cars with common characteristics using clustering based techniques. Figure \ref{arch} describes the system's high level architecture.

\section{Conclusion}
This work has introduced a full working HMM solution for connected cars cyber security with a unique approach handling the vehicle's collected data and a new temporal based detection technique. For each vehicle in the fleet, important data is collected using rule based engine. The collected data is tested (either in offline or online) against an HMM model trained on the vehicle's normal behavior. In the search for a smart detection technique, we built a regression model based on temporal features (e.g., time since drive start,consecutive arrival times,etc.) which predicted the expected Log-Likelihood at time $t$ and compared the result with the actual Log-Likelihood. Although the system performed very well even with static thresholds, the temporal detection technique proved its superiority and provided significantly
improved results.

Our system monitors not only the CAN bus, but rather monitors different vehicle's interfaces (OS, Network and CAN), and was proven to be highly capable to detect real life complicated anomalies which involves wide number of features from the different interfaces.

The deployment method we used to implement our solution is capable of monitoring car fleets. Furthermore, it allows us to train HMMs either for individuals or for groups of cars with common characteristics using a clustering preprocessing stage.

This work can serve as a basis for applying advanced analytics for identifying correlations between vehicles
as well as
in-progress attacks on vehicles using anomalies detected from other vehicles.

\bibliographystyle{IEEEtran}
\bibliography{references} 

\ifdefined\showAppendix

\appendices

\section{Events List} \label{events}

\begin{longtable}[c]{| p{3cm} | p{4cm} |}
\caption{Event types and their attributes\label{long}}\\
\hline
\bf{Event type} & \bf{Attributes} \\
\hline\hline
Login & location, velocity, source/destination packets\\
\hline
Door Unlocked & location, velocity\\
\hline
Door Opened & location, velocity\\
\hline
Door Closed & location, velocity\\
\hline
Fasten Seatbelt & location, velocity\\
\hline
Alarm Disarming & location, velocity\\
\hline
Ignition & location, velocity, source/destination packets\\
\hline
Release Seatbelt & location, velocity\\
\hline
Engine Stop & location, velocity\\
\hline
Door Locked & location, velocity, source/destination packets\\
\hline
Alarm Arming & location, velocity\\
\hline
Logout & location, velocity, source/destination packets\\
\hline
USB insert & location, velocity, source/destination packets\\
\hline
Authentication Process & location, velocity, source/destination packets\\
\hline
Running Exe File From USB & location, velocity, source/destination packets, file type\\
\hline
USB Extract & location, velocity, source/destination packets\\
\hline
Running App & location, velocity, source/destination packets\\
\hline
Open Flows & location, velocity, source/destination packets\\
\hline
Download App & location, velocity, source/destination packets, app name, path\\
\hline
List Of New Exec On The ECU & location, velocity, source/destination packets, app name, path\\
\hline
Network Usage & location, velocity, source/destination packets\\
\hline
Abnormal CAN behavior & location, velocity, source/destination packets\\
\hline
Abnormal NW behavior & location, velocity, source/destination packets\\
\hline
Abnormal OS behavior & location, velocity, source/destination packets\\
\hline
Main Router Login & location, velocity, source/destination packets\\
\hline
Start Download Firmware updates & location, velocity, source/destination packets\\
\hline
Finish Download Firmware updates & location, velocity, source/destination packets\\
\hline
Request Update & location, velocity, source/destination packets\\
\hline
Start firmware Update & location, velocity, source/destination packets\\
\hline
Finish firmware Update & location, velocity, source/destination packets\\
\hline
Main Router Logout & location, velocity, source/destination packets\\
\hline
Beacons & location, velocity, source/destination packets\\
\hline
OEM Communication & location, velocity, source/destination packets, OEM name\\
\hline
File Access & location, velocity, source/destination packets, file type, access type\\
\hline
Change In Data File Size & location, velocity, source/destination packets, file type, path\\
\hline
Map Process Started & location, velocity, source/destination packets\\
\hline
Loading Map & location, velocity, source/destination packets, path\\
\hline
Unknown Process Started & location, velocity, source/destination packets, path\\
\hline
Unknown Vendor Communication & location, velocity, source/destination packets, vendor name\\
\hline
GPS Access & location, velocity, source/destination packets\\
\hline
Open Ports & location, velocity, source/destination packets, ports list\\
\hline
Bluetooth Device Connected & location, velocity, source/destination packets, device id\\
\hline
Bluetooth Device Disconnected & location, velocity, source/destination packets, device id \\
\hline
\end{longtable}

\section{Stories} \label{Stories}
As described above, each story is a sequence of events (attributes could be changed in the different simulations and therefore we will specify only the important ones).

\textbf{Car Entry 1:} {\em  Door Unlocked, Door Opened, Door Closed, Fasten Seatbelt, Alarm Disarming, Ignition, Login}

\textbf{Car Entry 2:} {\em Door Unlocked, Door Opened, Door Closed, Login, Fasten Seatbelt, Release Seatbelt, Door Opened, Door Closed, Fasten Seatbelt, Alarm Disarming, Ignition}

\textbf{Car Exit:} {\em Engine Stop, Release Seatbelt, Door Opened, Door Closed, Door Locked, Alarm Arming, Logout}

\textbf{Drive Cancellation 1:} {\em Door Unlocked, Door Opened, Door Closed, Door Locked}

\textbf{Drive Cancellation 2:} {\em Door Unlocked, Door Opened, Door Closed, Alarm Disarming, Door Opened, Door Closed, Door Locked, Alarm Arming}

\textbf{Info System Upgrade:} {\em USB insert, Authentication Process, Running Exe File From USB, File Access , USB Extract}

\textbf{OTA Update:} {\em Main Router Login, Start Download FW, Finish Download FW, Request Update, Start FW Update, Finish FW Update, Main Router Logout}

\textbf{Play Music:} {\em Running App, Open flows, Network usage}

\textbf{Install App:} {\em Open Flows, Download App,  File access, Running App,List Of New Executables On The ECU}

\textbf{Download Map:} {\em OEM Communication, File Access, Change In Data File Size, Map Process Started, Loading Map}

\textbf{Music From USB 1:} {\em USB Insert, File Access, USB Extract} (the last event occurs at random time after file access)

\textbf{Music From USB 2:} {\em 1.	USB Insert, USB Extract, USB Insert, File Access, USB Extract} (the last event occurs at random time after file access)

\textbf{Music From USB 3:} {\em USB Insert, USB Extract, USB Insert, USB Extract} (the last event occurs at random time after file access)

\textbf{Music From mobile:} {\em Bluetooth Device Connected, File Access, Bluetooth Device Disconnected} (the last event occurs at random time after file access)

\textbf{Connected Device -– Mobile:} {\em Bluetooth Device Connected (id: known/unknown id)}

\textbf{Open flows -- GPS application:} {\em Open Flows (TargetIP: 130.211.9.172. Priority: low(2))}

\textbf{Open flows -- Weather application:} {\em Open Flows (TargetIP: 46.228.47.115. Priority: low(2))}

\textbf{GPS:} {\em GPS access}

\textbf{Open Ports:} {\em Open Ports}

\fi

\end{document}